\documentstyle[prl,aps,epsf]{revtex}
\begin{document} \draft 
\twocolumn[\hsize\textwidth\columnwidth\hsize\csname @twocolumnfalse\endcsname
\title{Peltier Current Lead Experiments with a Thermoelectric Semiconductor near 77 K}
\author{Satarou YAMAGUCHI, Hiroaki NAKAMURA, Kazuaki IKEDA$^a$,\\
Tokushi SAKURAI$^b$, Ikushi YOSHIDA$^b$, Seiichi TANUMA$^b$,\\
Satoshi TOBISE$^c$, Kunihito KOUMOTO$^c$,\\
Haruhiko OKUMURA$^d$} 
\address{National Institute for Fusion Science (NIFS), Oroshi-Cho, Toki-City, Gifu-Prefecture, 509-52, Japan, Phone\&Fax : +81-52-789-4538, E-mail:yamax@ysl.nifs.ac.jp,hiroaki@rouge.nifs.ac.jp,\\ 
http://rouge.nifs.ac.jp/$\tilde{\ \ }$hiroaki/index.html}
\address{$^a$Department of Fusion Science, The Graduate University for 
Advanced Studies (GUAS),
322-6, Oroshi-Cho, Toki-City, Gifu-Prefecture, 509-52, Japan}
\address{$^b$Iwaki Meisei Univ.,Chuodai-Iino 5-5-1, Iwaki 970, Japan}
\address{$^c$Nagoya Univ., Furo-cho, Chikusa-ku, Nagoya 464-01, Japan.}
\address{$^d$Matsusaka Univ., Kubo-cho 1846, Matsusaka 515, Japan.}
\date{\mbox{This paper was presented in 'OD1-7', MT-15, Beijin, China (1997).} } 
\maketitle

\begin{abstract}
Peltier current lead was proposed to reduce heat leak from the current lead. 
The temperature of the hot side of semiconductors was kept to be room temperature and the liquid nitrogen was used to cool the system in 
the experiment.  The experiment confirmed the principle of the Peltier 
current lead, and the reduction of the heat leak is calculated to be 30 
\% for the liquid helium system and 40 \% for the liquid nitrogen system.  
We also proposed a new current lead system which is composed of semiconductors 
and high temperature superconducting material (HTS).  This idea bases on
the functionally gradient material (FGM), and the HTS is connected 
to the semiconductor directly. The temperature of the hot side of 
semiconductor is kept to be the liquid nitrogen temperature, 
the temperature of HTS can be expected to be lower than 77 K. Therefore, 
we can expect high current capacity of the HTS and/or high stability 
of the HTS. We use BiSb as a N-type semiconductor and BiTe as a P-type 
semiconductor in the experiment, and the temperature of the cold 
side of the semiconductor is 73 K in this experiment.
\end{abstract}
\vskip2pc] \narrowtext
 
\section{INTRODUCTION}

Peltier current lead was proposed, and the P-type and N-type semiconductors made of BiTe were kept at the hot side of room temperature side in the demonstration experiment.  The temperature difference between the high and low temperature sides of the semiconductors was around 100 K, which confirmed the principle of the Peltier current lead \cite{ref1}.  A superconducting magnet system includes a room temperature power supply and a low temperature magnet, which creates a temperature difference in the electric circuit.  This can be dealt with by using an thermoelectric element take the Peltier current lead. Analyzing and numerical calculations gave theoretical results that confirmed the original experimental findings \cite{ref2}.  While liquid nitrogen was used as a coolant in the experiment, resulting in a  40 \% reduction in the heat leak from the current lead, the estimated reduction in a liquid helium superconducting system would be about 30 \%.  If the system did not use a coolant, the Peltier current lead would be even more effective.

Electric power consumption increases when the semiconductors are inserted, but the efficiency of the electric power is improved ten times because the heat leak is reduced and the efficiency of refrigerator is low. Since A cooling system is not cheap, the Peltier current lead can reduce not only the running costs but also the instrument costs of the system. It has an active element, i.e. its thermoelectric effect. Moreover, since the thermal conductivity of BiTe is only 0.3 to 0.5 \% that of copper  around room temperature, so even if the current of the current lead is zero, the heat leak is reduced. This is a passive effect.

	A high temperature superconductor (HTS) has recently been used in so-called HTS current leads \cite{ref3}.  It is a passive element, i.e. the thermal conductivity of HTS material is currently about  0.1 \% that of the copper below 77 K, reducing the occurrence of  the heat leak from 77 K to the temperature of liquid helium. The conductivity of HTS is of the same order as that of the semiconductor in this temperature range.  While BiTe is a good semiconductor material near room temperature,  i.e. a BiTe semiconductor offers low thermal conductivity, high electrical conductivity and high thermoelectric power, other semiconductors and/or multistage elements are needed because the low temperature side of the semiconductor reaches around 200 K in the maximum temperature difference operation.  BiSb is a good N-type semiconductor material in low temperatures and it is connected with HTS \cite{ref4}.  BiSb also has a magnetic field effect, so its performance improves \cite{ref4} when a magnetic field is applied.

	A number of  HTS materials has been developed and studied and Bi-2223 is one of the best HTS material for the current lead.  All materials of the semiconductor and HTS are, therefore, based on Bi alloy designed for temperature from room temperature to 77 K. Yamaguchi et al \cite{ref5} proposed a new current lead concept based on functionally gradient material (FGM). Low temperature experiments were needed to realize this concept. We conducted and analyzed these low temperature experiments, where the hot temperature side of the semiconductors was set at the temperature of liquid nitrogen.

\section{EXPERIMENTAL SET UP}
	Figure 1 shows the experimental device.  We used liquid nitrogen to keep the temperature constant, on the hot temperature side of the semiconductors which were connected to the immersed copper plate electrodes by low temperature solder.  The semiconductors were set in a vacuum vessel with a vacuum of 
$2.0 \times 10^{-6}$ 
torr for thermal insulation. The N-type semiconductor, made of BiSb, 
was 12 mm diameter and 10 mm length, the P-type semiconductor, 
a made of BiTe, and has a cross section of $13 \times 13$ mm 
and the length of 10 mm. The transport coefficients of these semiconductors and the copper near 77 K are listed in Table 1, 
$\alpha$ the thermoelectric power, 
$\kappa$ the thermal conductivity and $\eta$ the electrical resistivity.  
The thermal conductivity of copper is increases with a decrease in temperature and reaches about 1200 W/m/K near 20 K. The thermal conductivity of the semiconductors decreases with the decreasing temperature but no more than 30 \%. The electrical resistivity of copper decreases with a decrease in temperature and reaches saturation of 0.0002 to 0.0001 $ \mu \Omega$m below 20 K, 
while the resistivity of the semiconductors increases with a decrease in temperature without reaching saturation. 
Since the low temperature sides of the two semiconductors are connected 
by a flexible copper cable, its temperature is lower than that of the 
liquid nitrogen. The performances and the cold temperatures of BiTe and 
BiSb are different from each other, but this difference is not so 
big because the thermal conductivity of the copper cable is high.

\begin{figure}
\epsfxsize=8cm \epsffile{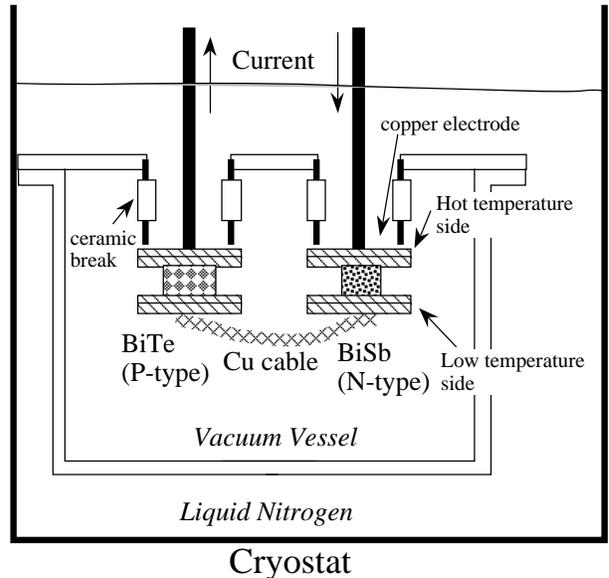}
\caption{Experimental device. Semiconductors are set in a vacuum vessel for thermal insulation. The temperature of the hot side of the semiconductors is kept at the temperature of the liquid nitrogen.}
\label{fig.1}
\end{figure}

 Temperatures were measured by Type C thermocouples, directly connected to the electrodes by the low temperature solder, and voltages were measured by the thermocouples cable.   Electrical current is controlled by the power supply with an accuracy of 3 digits.

\section{EXPERIMENTAL RESULT AND CALCULATION}
	Figure 2 shows the voltages of the semiconductors at different currents.  The voltages were fairly proportional to the currents. The semiconductor resistance accounted for the main part of their inclines. Others, 
small factors were the electrode contact resistance measured at 3 to 6 $ \mu \Omega$,
the copper electrode resistivity is quite small (see Table 1) and 
the thermoelectric voltages can be estimated from Table 1. 
These estimations indicate that the main part of the inclines come 
from the resistance of the semiconductors. 
The estimated semiconductors resistances was 
132 $ \mu\Omega$  for BiTe and 83 $ \mu\Omega$  for BiSb.

	Figure 3 shows the measured temperatures of the semiconductors at 
different currents. Measured in the center of the copper plate electrodes, 
the lowest temperature of BiTe was occurred at a current of 30 to 40 A, 
while the lowest temperature of BiSb did not materialize over the current 
range of our experiment. The temperature difference between the hot and 
the cold side was greater with BiSb than with BiTe, which makes BiSb a
better material for the temperature range of our experiment. 

\begin{figure}
\epsfxsize=6cm \epsffile{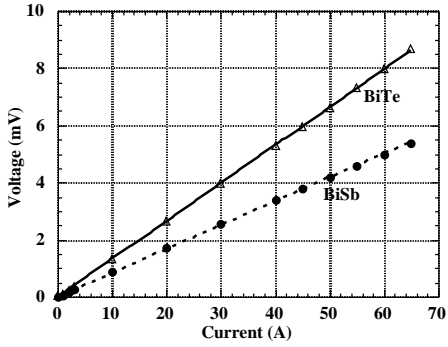}
\caption{Currents and voltages of the semiconductors.}
\label{fig.2}
\end{figure}

\begin{figure}
\epsfxsize=6cm \epsffile{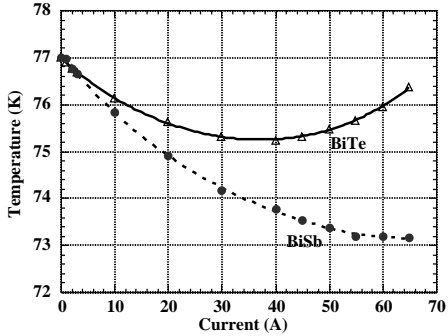}
\caption{Current versus the low temperatures of semiconductors in the experiment.}
\label{fig.3}
\end{figure}

Theoretically, the optimum current $I_{\rm opt},$ i.e.  
the maximum temperature difference between two 
sides of a semiconductor, is given by
\begin{eqnarray}
I_{\rm opt} &=& \alpha \frac{T_{\rm c}}{r_{\rm e}} \nonumber \\
 &=& \frac{K_{\rm e} }{\alpha} 
\left\{ 
  \left[ 1+ 2 Z\left( T_{\rm h} + \frac{q_{\rm c} }{ K_{\rm e} } \right)
  \right]^{0.5} 
\right\}, \label{eq.1}
\end{eqnarray}
where $T_{\rm c}$ is the temperature of the cold side i.e., $T_{\rm h}$
is the temperature of the hot side, almost 77 K in our experiment; 
and $q_{\rm c}$ is the heat flux to the cold side. 
The other parameters are defined by
\begin{eqnarray}
K_{\rm e} &=& \kappa \frac{S}{l} , \label{eq.2} \\
r_{\rm e} &=& \eta \frac{l}{S}, \label{eq.3} \\
Z	&=& \frac{\alpha^2}{\kappa \eta}, \label{eq.4}
\end{eqnarray}
where $K_{\rm e}$ is the thermal conductance, 
$r_{\rm e}$ is the electrical resistance, 
$Z$ is the figure of merit,  
$S$ is the cross section area of the semiconductor, and 
$l$ is its length.

The optimum current was calculated to use Table 1, 
and was about 43 A for the BiTe and 140 A for the BiSb.  
The temperature of the cold side of semiconductor 
is a function of current and given by 
\begin{equation}
T_{\rm c} = \frac{1}{\alpha I + K_{\rm e}} 
\left(
   \frac{1}{2} r_{\rm e} I^2 + K_{\rm e} T_{\rm h} + q_{\rm c} 
\right),	\label{eq.5}
\end{equation}
where $I$ is the current of the semiconductor.

In order to calculate the temperature of the cold side, we assumed the followings;
\begin{eqnarray}
T_{\rm h} &=& 77 {\rm K}, \label{eq.6} \\
q_{\rm c} &=& R_{\rm int} I^2,	\label{eq.7}
\end{eqnarray}
where $R_{\rm int}$ is the total resistance of the cold side of the semiconductors.
\begin{table}
\epsfxsize=8cm \epsffile{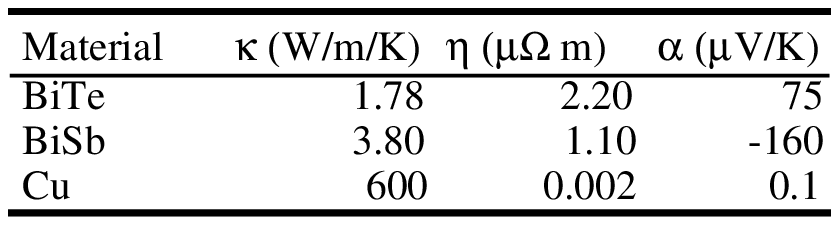}
\caption{Transport coefficients of BiTe, BiSb and copper near 77 K.}
\label{tab.1}
\end{table}

We measured the voltage of the copper cable in the experiment, and the result is Fig. 4.  Rint  could be estimated as 38.8 $\mu\Omega$. Having determined the other parameters of Eq. (5) based on the coefficients of Table 1, we could finally calculate the temperature of the cold side of the semiconductors at different currents.  See Fig. 5. The lowest temperature of BiTe was obtained at a current of 30 A, while the lowest temperature of BiSb had not been obtained at the upper end of the current range of our calculation.  While the graphs of the measured temperatures in Fig. 3 are very much like the graphs of the calculated temperature in Fig. 5, however,  the absolute temperature values are different.

One reason for this difference is the copper cable connection between two semiconductors . Its thermal conductivity is high, so the BiSb would pump out the heat flux from the BiTe side through the copper cable. Because of this, the assumption of Eq. (7) is not suitable or the values of the $R_{\rm int}$
is not the same for the two semiconductors. The transport coefficients (thermoelectric power, electrical resistivity and thermal conductivity in this study) 
are temperature-dependent, so we should analyze a nonlinear equation.

\begin{figure}
\epsfxsize=6cm \epsffile{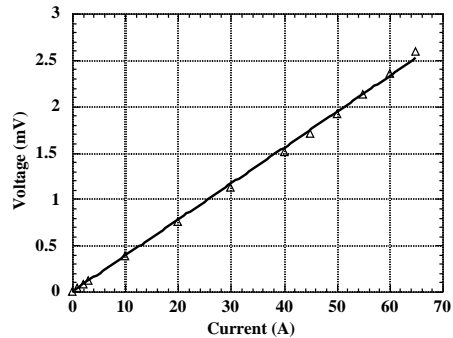}
\caption{Currents and voltage of the copper cable.}
\label{fig.4}
\end{figure}

\begin{figure}
\epsfxsize=6cm \epsffile{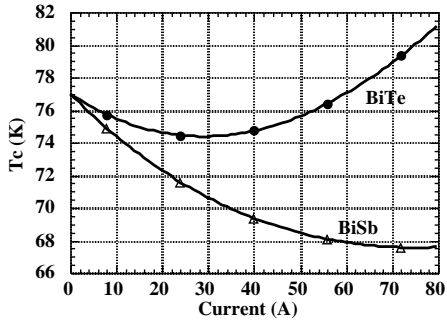}
\caption{Currents and calculated temperature on the cold side of  the semiconductors.}
\label{fig.5}
\end{figure}

\section{DISCUSSION AND CONCLUSION}

Both the P-type BiTe and N-type BiSb semiconductors work below the temperature of liquid nitrogen, but their performances  are different. In an actual current lead, the semiconductors are individually connected to the HTS component, and thus thermally insulated.  This will simulate our experiment and the next experiments by inserting the HTS component into the cold side of the two semiconductors. With the thermal conductivity of HTS material only absent 0.1 \% that of copper below 77 K, it will be easy to thermally insulated the two semiconductors even if we use only a short HTS component. We will also improve the shape of the semiconductors to take into account that the optimum currents of BiTe and BiSb are so different from each other, and that this reduces the overall performance of the Peltier current lead.

The BiTe and BiSb materials were not  optimized. BiTe is would have been most efficient around room temperature, and the BiSb sample was hand made with less than optimal for the doping material and density. Having used these multi-crystal materials, we should also consider using single-crystal material.

On the other hand, multi-crystal materials have been described with higher performance atomic ratios than Bi and Sb  (88 \% and 12 \%). These materials have a magnetic field effect, and the magnetic field effect should improve overall performance.

With the high temperature on the hot side of the HTS material being brought down from 77 K by the semiconductors, its current capacity is increased and/or its stability improved. Using the parameters from the  references, the temperature drop was calculated to have been 7 to 12 K for BiTe and BiSb at  optimal currents.  The drop in temperature is more effective to use HTS. 

\acknowledgments
	The authors would like to thank Prof. K. Kuroda, Nagoya University for his constructive criticism, and Prof. A. Iiyoshi, director general of the National Institute for Fusion Science, for his support through this study.

\end{document}